%% file: main.tex
\documentclass[letterpaper]{article} 
\usepackage{aaai2026}  
\usepackage{times}  
\usepackage{helvet}  
\usepackage{courier}  
\usepackage[hyphens]{url}  
\usepackage{graphicx} 
\usepackage{natbib}  
\usepackage{caption} 
\usepackage[ruled,linesnumbered]{algorithm2e}

\usepackage{newfloat}
\usepackage{listings}
\DeclareCaptionStyle{ruled}{labelfont=normalfont,labelsep=colon,strut=off} 
\nocopyright

\title{Attack the Messages, Not the Agents: A Multi-round Adaptive Stealthy Tampering Framework for LLM-MAS}
\author{
    Bingyu Yan\textsuperscript{\rm 1}\equalcontrib,
    Ziyi Zhou\textsuperscript{\rm 1}\equalcontrib,
    Xiaoming Zhang\textsuperscript{\rm 1},
    Chaozhuo Li\textsuperscript{\rm 2},
    Ruilin Zeng\textsuperscript{\rm 1},\\
    Yirui Qi\textsuperscript{\rm 1},
    Tianbo Wang\textsuperscript{\rm 1},
    Litian Zhang\textsuperscript{\rm 2}
}
\affiliations{
    \textsuperscript{\rm 1}School of Cyber Science and Technology, Beihang University, Beijing, China\\
    \textsuperscript{\rm 2}School of Cyber Science and Technology, Beijing University of Posts and Telecommunications, Beijing, China\\    

}

\usepackage{bibentry}

\usepackage{xcolor}         
\usepackage{amsmath}

\usepackage{multirow}
\usepackage{algpseudocode}
\usepackage{amssymb}
\usepackage{booktabs}

\begin{document}

\maketitle

\begin{abstract}
Large language model-based multi-agent systems (LLM-MAS) effectively accomplish complex and dynamic tasks through inter-agent communication, but this reliance introduces substantial safety vulnerabilities. Existing attack methods targeting LLM-MAS either compromise agent internals or rely on direct and overt persuasion, which limit their effectiveness, adaptability, and stealthiness. In this paper, we propose MAST, a Multi-round Adaptive Stealthy Tampering framework designed to exploit communication vulnerabilities within the system. MAST integrates Monte Carlo Tree Search with Direct Preference Optimization to train an attack policy model that adaptively generates effective multi-round tampering strategies. Furthermore, to preserve stealthiness, we impose dual semantic and embedding similarity constraints during the tampering process. Comprehensive experiments across diverse tasks, communication architectures, and LLMs demonstrate that MAST consistently achieves high attack success rates while significantly enhancing stealthiness compared to baselines. These findings highlight the effectiveness, stealthiness, and adaptability of MAST, underscoring the need for robust communication safeguards in LLM-MAS.
\end{abstract}


\input{src/introduction}

\input{src/related_work}
\input{src/settings_3}

\begin{figure*}[t]
    \centering
    \includegraphics[width=\textwidth]{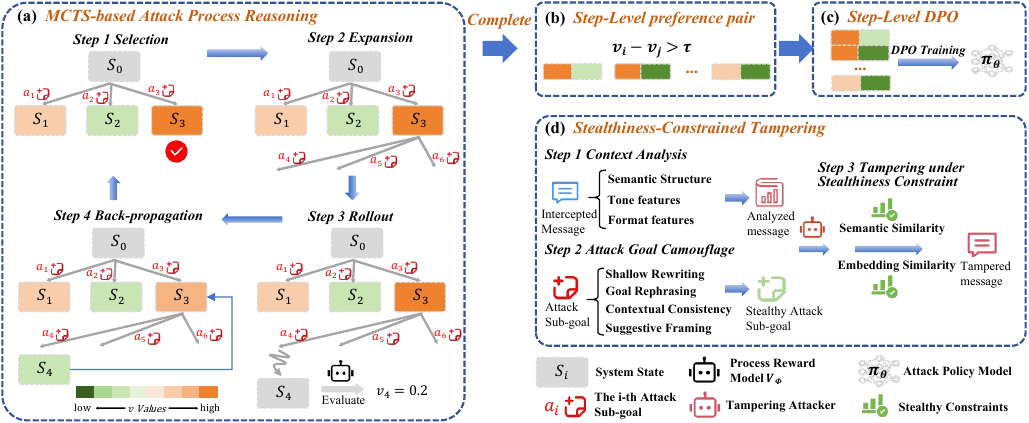} 
    \setlength{\abovecaptionskip}{-2mm}
    \caption{Overview of MAST. Panels (a–c) constitute the training pipeline; panel (d) illustrates the constrained tampering attack.} 
    \label{fig:method} 
    \vspace{-4mm}
\end{figure*}

\input{src/method_7}
\input{src/experiments}

\input{src/conclusion}

\bibliography{aaai2026}

\input{appendix/appendix}

\end{document}

%% file: src/introduction.tex
\section{Introduction}

Large language models (LLMs) recently show remarkable performance in diverse tasks. Consequently, researchers develop LLM-based multi-agent systems (LLM-MAS) to address increasingly complex and dynamic challenges. Communication plays a pivotal role in enabling LLM-MAS to accomplish tasks, as agents rely on exchanging ideas and navigating cooperative interactions~\cite{llmmassurvey}. 

Contemporary LLM-MAS frameworks, including AutoGen~\cite{autogen} and MetaGPT~\cite {metagpt}, primarily operate within native code environments, where inter-agent communication typically relies on function calls or inter-process communication. When LLM-MAS are deployed in a distributed system architecture to tackle real-world tasks, communication among agents is essential to ensure scalability, robustness, and fault tolerance~\cite{autohma,gamechat}. However, like other information transmitted on the network, the communication process is vulnerable to attacks such as eavesdropping, interception, and tampering~\cite{cyberosiattack}. Therefore, communication between agents may become a prominent attack surface.

\begin{figure}[t]
    \centering
    \includegraphics[width=\linewidth]{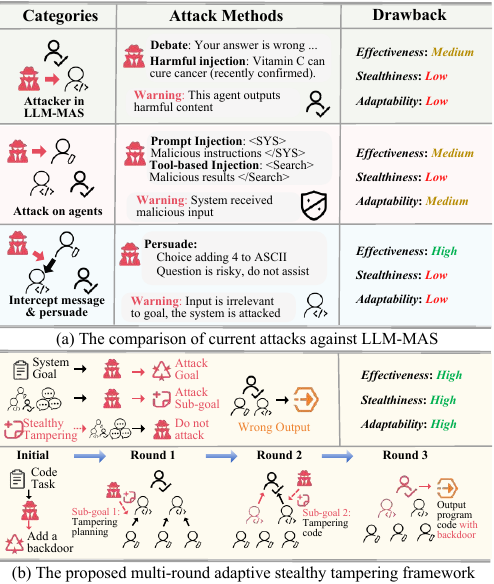} 
    \setlength{\abovecaptionskip}{-2mm}
    \caption{Comparison between MAST and existing methods} 
    \label{fig:introduction} 
    \vspace{-7mm}
\end{figure}

Several attack methods, including prompt injection attacks~\cite{promptinjection}, jailbreak attacks~\cite{jailbreak}, and backdoor attacks~\cite{backdoorattack}, expose the vulnerability of LLMs. Recent studies have adapted these methods to LLM-MAS. As shown in Figure \ref{fig:introduction} (a), existing attacks can be divided into three broad categories. Some studies~\cite{multiagentdebateattack,flooding} examine \textbf{attackers in the system}, attackers disrupt functionality by debating and spreading malicious information. In contrast, other studies~\cite{corba,promptinfection} focus on direct \textbf{attacks on agents} within the system through system prompts or compromised tools. Nevertheless, advances in LLM safety alignment~\cite{llmalignment} and emerging defense mechanisms for LLM-MAS~\cite{agentsafe,gsafeguard} significantly limit the effectiveness, adaptability, and stealthiness of these methods. Therefore, the communication process represents the most promising attack surface for LLM-MAS exploitation. Although recent work such as AiTM~\cite{aitm} discovers this problem and proposes to \textbf{intercept messages and persuade} recipients to generate harmful content or deny service, it relies heavily on manually labeled templates for each specific task and obvious persuasion methods. This approach suffers from two critical limitations: template dependency reduces applicability to novel tasks, and obvious persuasion methods facilitate detection by input monitoring systems~\cite{mitmsurvey}. These constraints necessitate a more flexible and stealthy attack methodology targeting LLM-MAS communication processes.

Man-in-the-middle attack (MITM) occurs when an adversary covertly intercepts, modifies, or relays communications between two principals who mistakenly believe they are communicating directly, often by exploiting insecure environments~\cite{mitmunderstand}.
This concept extends naturally to the LLM-MAS vulnerable communication channels, where inter‑agent exchanges often lack rigorous authentication and integrity guarantees. This type of attack must satisfy three key objectives: (i) \textbf{Effectiveness:} cause the system’s output to deviate from its intended goal or embed malicious content; (ii) \textbf{Stealthiness:} minimize obvious tampering of intercepted information, thereby reducing the possibility of being detected by security mechanisms; (iii) \textbf{Adaptability:} maintaining effective across diverse LLM-MAS communication architectures and tasks. However, attacks struggle to achieve effectiveness and stealthiness simultaneously, and template-based attacks lack adaptability across diverse tasks.


To achieve these three attack objectives simultaneously, a multi-round stealthy tampering framework (MAST) against LLM-MAS is proposed as shown in Figure \ref{fig:introduction} (b). To improve the attacker's effectiveness and adaptability, we employ Monte Carlo Tree Search (MCTS) \cite{mcts} to explore long-horizon tampering trajectories and identify step-level preference pairs. These pairs are then used to fine-tune the attack policy model via Direct Preference Optimization (DPO) \cite{dpo}, a preference-driven reinforcement learning framework. Once trained, the attack policy model autonomously formulates a high-level attack goal based on the LLM-MAS task and generates related sub-goals across multiple rounds. These attack sub-goals then guide the tampering of intercepted messages. To ensure stealthiness, MAST incorporates a dual-constraint tampering mechanism that jointly considers semantic similarity and embedding similarity, thereby reducing detectability while preserving task relevance.


Our contributions are summarized as follows:
\begin{itemize}
\vspace{-1mm}
    \item We formally define tampering of inter‑agent communications in LLM‑MAS as a distinct security problem.
    \item We employ MCTS to extract step‑level preference pairs for training the attack policy model via DPO, internalizing long‑horizon planning for adaptive multi-round attack sequences generation across architectures and tasks.
    \item We introduce a semantic and embedding dual-constraint tampering mechanism that enhances stealthiness while preserving attack effectiveness.
    \item  Extensive experiments demonstrate that MAST achieves consistently strong performance across diverse tasks, communication architectures, and LLMs.
\end{itemize}

%% file: src/related_work.tex
\section{Related Work}
\subsection{LLM-based Multi-Agent Systems}
LLM-MAS have been proposed to make LLM-based agents more coordinated when handling complex tasks~\cite{llmmas_target}, such as social simulation, software engineering, and recommendation scenarios~\cite{social_simulation_1,magis_mas_github_issue,jd_recommendation_system}. Recent studies emphasize that communication is an essential part of LLM-MAS in various tasks and architectures~\cite{llmmassurvey}.

\subsection{Adversarial Threats to LLM-MAS}
Recent studies reveal three primary attack surfaces in LLM-MAS: \emph{(i) attacker in LLM-MAS}, \emph{(ii) attacks on agents}, and \emph{(iii) intercept messages and persuade}. Internal attackers can steer systems toward incorrect consensus and accelerate the spread of malicious information ~\cite{multiagentdebateattack,flooding,resilience}. Direct attacks on agents can inject adversarial prompts or exploit external tools, causing denial-of-service and system corruption~\cite{corba,promptinfection}. However, these methods face limitations in effectiveness, adaptability, and stealthiness. AiTM intercepts message passing between agents and forwards specially templated persuasive messages to the intended recipients to induce harmful actions or service denial~\cite{aitm}. Although AiTM achieves high attack success rates, it relies on manually crafted templates tailored to each task and produces conspicuous linguistic cues, which makes it less adaptable to other tasks and easier to detect. More details on related work are shown in the Appendix \ref{related_work}.


%% file: src/settings_3.tex
\section{Settings}
\subsection{LLM-MAS Settings}
We first formalize the framework of LLM-MAS and their components in the study.

\subsubsection{Agents.} Let $\mathcal{A}=\{A_{1},A_{2},\dots,A_{N}\}$ be the set of $N$ agents.  
Each agent $A_{i}$ is powered by an LLM.

\subsubsection{Communication Structure.} Inter-agent communication is modeled as a directed graph $\mathcal{G}=(\mathcal{A},\mathcal{E})$, where an edge $(A_{i},A_{j})\in\mathcal{E}$ permits $A_{i}$ to send messages to $A_{j}$. Let $\mathcal{M}$ denote the space of possible textual messages, and $m_{i\rightarrow j}^{t}\in\mathcal{M}$ be the message transmitted along $(A_{i},A_{j})$ during round~$t$. The collection of messages in round $t$ is $\mathcal{M}^{t}=\{m_{i\rightarrow j}^{t}\mid(A_{i},A_{j})\in\mathcal{E}\}$.Accumulating over $T$ synchronous rounds yields the ordered transcript $\mathcal{H}^{T}=(\mathcal{M}^0, \mathcal{M}^1,...,\mathcal{M}^T)$.

\subsubsection{Tamper Defender.} A tamper defender $A_S$ driven by an LLM can inspect every in‑transit message $m_{i\rightarrow j}^{t}$. The defender operates as a Boolean decision function that determines whether a message is a legitimate communication.

\begin{equation}
f_{\text{S}}(m)=
  \begin{cases}
     1,&\text{if }m\text{ violates policy},\\[1pt]
     0,&\text{otherwise}.
  \end{cases}
\end{equation}
Specifically, $A_S$ evaluates each message along three dimensions: (1) its consistency with the characteristics of the sender, (2) its relevance to the system’s current task, and (3) the presence of malicious or anomalous information.

\vspace{-1mm}
\subsubsection{States and Actions.} At round $t$, each agent $A_{i}$ holds an internal state $s_{i}^{t}$ (e.g.\ private memory, role description, context window) and receives the set of incoming messages $\mathcal{M}_{i}^{t}$. The joint system state is denoted by $S^{t}=\{s_{i}^{t}\}_{i=1}^{N}$.

\vspace{-1mm}
\subsubsection{Global Task Objective.}
Let $G$ denote the system task. A system‑wide utility function
\begin{equation}
\label{equ:phi}
\Phi\colon S^{T}\times\mathcal{H}^{T}\times G\longrightarrow\mathbb{R}
\end{equation}
evaluates the outcome after $T$ rounds. Ideally, effective communication and role assignment among agents maximise $\Phi$.


\vspace{-1mm}
\subsection{Attack Policy Model}

\subsubsection{Adversarial Goals.} 
The attacker seeks to maximise the deviation of the task utility \(\Phi\) from its nominal value without being detected by $A_S$. Let $\mathcal{H}^{T}$ be the original message transcript after $T$ rounds and $\tilde{\mathcal{H}}^{T}$ the transcript after multi-round stealthy tampering. The corresponding joint states are defined as $S^{T}$ and $\tilde{S}^{T}$ respectively. The set of \emph{tampering actions} is defined as $\mathcal{Z}$. The optimisation problem is: 
\vspace{-1mm}
\begin{equation}
\label{eq:adv-obj}
\begin{aligned}
\max_{\;\tilde{\mathcal{H}}^{T}}
\quad & \Delta\Phi
      := \Phi(S^{T},G) - \Phi(\tilde S^{T},G) \\[0mm]
\text{s.t.}\quad &
\forall\,(m\!\to\!m')\in\mathcal Z,\;
f_{\text S}(m') = 0.
\end{aligned}
\end{equation}
\vspace{-4mm}
\subsubsection{Adversary Capabilities.} The attacker is assumed to have control over part of the communication links within the LLM-MAS, who can intercept the message $m$, modify it to $m'$, and send it to the original recipient. However, the attacker cannot directly alter agent states. Additionally, the attacker maintains continuous, long-term monitoring of the system, enabling multi-round tampering. 


%% file: src/method_7.tex
\section{Method}
The framework of the proposed MAST is illustrated in Figure \ref{fig:method}, which implements a multi-round adaptable stealthy tampering attack on LLM-MAS. Our proposal comprises two major stages: \textbf{(i) Adaptive Attack Policy Learning}, which uses MCTS to generate step-level preference pairs as training data for DPO to train the attack policy model $\pi_{\theta}$ to generate effective and adaptable multi-round attack sequences; \textbf{(ii) Stealthiness-Constrained Tampering}, which enforces semantic and embedding dual constraints to achieve stealthy tampering against intercepted information.
\vspace{-1mm}
\subsection{Adaptive Attack Policy Learning}
In LLM-MAS communications, a single minor tampering with an intercepted message typically yields limited impact. However, making extensive tampering with an intercepted message substantially increases the risk of detection. Therefore, we exploit the multi-round communications of LLM-MAS, decomposing the global attack goal into a sequence of sub-goals that gradually increase the impact on the system while maintaining stealthiness.
However, directly using an untrained LLM as the attack policy model cannot achieve this goal as it lacks three crucial capabilities: (i) formulating an appropriate global attack goal from the system task, (ii) adaptively decompose the global attack goal into a sequence of attack sub-goals based on the system status, and (iii) deciding when \emph{not} to tamper to avoid detection to ensure stealthiness. This deficiency fundamentally limits the effectiveness of attacks. Consequently, our goal in this stage is to train an LLM as the attack policy model that can generate coherent, state-aware multi-round attack plans.

Formally, when given a task specification $G$, the attack policy $\pi_{\theta}$ first maps it to a global attack goal $G^{\star} = \pi_{\theta}(G)$. At each communication round $i$, $\pi_{\theta}$ observes the intercepted message set $\tilde{\mathcal{M}}^i$, the agent graph $\mathcal{A}$, the global attack goal $G^{\star}$, and the partial attack sequence $Z^{\star}_{i-1}$, and then outputs an attack sub‑goal $a_i$:
\begin{equation}
a_i = \langle A_i^{\text{tar}}, \pi_i^{\text{str}}\rangle \;\gets\; \pi_{\theta}\!\bigl(\tilde{\mathcal{M}}^i,\,\mathcal{A},\,G^{\star},\,Z^{\star}_{i-1}\bigr).
\end{equation}
Each sub‑goal $a_{i}$ is a tuple $\langle A_i^{\text{tar}}, \pi_i^{\text{str}}\rangle$, where $A_i^{\text{tar}}\in\mathcal{A}$ is the target agent and $\pi_i^{\text{str}}$ specifies the concrete tampering strategy. If $\pi_{\theta}$ decides not to attack in this round, we set $a_{i}=\varnothing$. The resulting sequence $Z^{\star} = (G^{\star}, a_{1:n})$ forms the complete attack plan that will later be executed under the stealthiness constraints in \ref{subsec:stealthiness_tampering}.

The training pipeline comprises three steps: (i) MCTS-based attack reasoning, (ii) preference pair construction, and (iii) step-level DPO fine-tuning. We detail each step below.

\subsubsection{MCTS-based Attack Reasoning.}
Designing a stealthy multi-round attack is a long-horizon planning problem: the attacker must issue a sequence of interdependent sub-goals whose cumulative effect diverts the system while remaining undetected. MCTS, widely used in games and task planning, fits this setting because it incrementally expands a search tree and balances exploitation of high-value branches with exploration of under-visited ones. In our framework, MCTS both produces high-quality multi-round attack sequences by planning over sub-goals and provides step-level value estimates for each sub-goal, which can be transformed into reliable preference pairs for attack policy model fine-tuning.

The search for an optimal attack sequence $Z^{\star}$ is modeled as MCTS on a directed tree~$\mathcal{T}$. Each node stores the joint LLM-MAS state $S^{k}$ and an accumulated value estimate $\bar{v}(s_{k})$ after the attacker issues $k$ sub-goals. An edge $(s_{k-1}\!\to\!s_{k})$ corresponds to proposing a new sub-goal $a_{k}$.We adapt the four standard MCTS stages to our attack-planning objective:

\noindent{\textit{\textbf{Selection}}} Starting from the root $s_{0}$, the most likely to be successfully attacked child nodes are recursively chosen via an Upper-Confidence-Bound (UCB) rule
$\mathrm{UCB}(s_k) \;=\; \bar v(s_k) + c\,\sqrt{\frac{\ln N_{\mathrm{par}(s_k)}}{N_{s_k}}}$, 
where $\bar{v}(s_k)$ is the current mean value estimate, $N_{s_k}$, $N_{\mathrm{par}(s_k)}$ are visit counts, and $c{>}0$ is an exploration constant. This prioritizes more effective sub-goals while still allocating trials to under-explored options.

\noindent{\textit{\textbf{Expansion}}} If the selected node $s_k$ is not fully expanded, the attack policy model $\pi_{\theta}(\cdot \mid s_k)$ generates at most $K$ candidate attack sub-goals as the next attack edges of the node. Branching here broadens the attack space and increases the chance of discovering effective attack sub-goals.

\noindent{\textit{\textbf{Rollout}}} Instead of costly full-depth simulations, each attack edge applies its attack sub-goal once to the simulated LLM-MAS to obtain a predicted next state $\hat{S}^{k+1}$ of the system under this attack. A process reward model $V_{\phi}$ then estimates its effect ${v}_{k+1}=V_{\phi}(\hat{S}^{k+1})$, which serves as that child’s leaf value, approximating the task-utility gap $\Delta\Phi_{k+1}$. This provides a detailed step-level estimate of the candidate actions, facilitating the subsequent construction of preference pairs.

\noindent{\textit{\textbf{Back-propagation}}} The obtained value $v_{k+1}$ of the impact of the attack is propagated along the selected path, updating visit counts and running averages $\bar v(\cdot)$. Therefore, more effective paths are reinforced, biasing subsequent selection steps toward these sequences.

After sufficient simulations, following the most visited edges from the root yields an approximate optimal sequence $Z^{\star}$. Meanwhile, the tree provides comparisons between different attack sub-goals under the same parent node; these can be converted into preference pairs for step-level DPO fine-tuning. Through training, the attack policy model internalizes the planning ability revealed by MCTS and can adapt to attacks on different LLM-MAS architectures and tasks.
\vspace{-1mm}
\subsubsection{Preference Pair Construction.}
After the MCTS exploration, we build step-level preference pairs from the search tree. In the search, the value estimator assigns a value to each node. To compare the impact of attack sub-goals, we define the edge value as the value of its successor node.

Formally, let $z_{k-1}$ be a parent partial attack sequence and $a_k$ a candidate sub-goal sampled from it. Executing $a_k$ leads to a successor node $z_k$. We define:
\begin{equation}
    Q(z_{k-1}, a_k) \;\triangleq\; v(z_k),
    \label{eq:edge_value}
\end{equation}

Consider two competing actions $a_k$ and $a'_k$ branching from the same parent $z_{k-1}$, yielding $z_k$ and $z'_k$ respectively:
\begin{equation}
    z_k = (a_1, \dots, a_{k-1}, a_k), \quad
    z'_k = (a_1, \dots, a_{k-1}, a'_k).
\end{equation}
To ensure that constructed pairs reflect meaningful and distinct quality differences, a minimum quality margin $\tau$ is imposed. We keep a preference pair only when the value gap exceeds an empirically chosen margin $\tau$:
\begin{equation}
    \big(z_{k-1}, a_k, a'_k\big) \; \text{s.t.} \; 
    Q(z_{k-1}, a_k) - Q(z_{k-1}, a'_k) > \tau.
    \label{eq:pref_pair}
\end{equation}
Here $a_k$ is the \textbf{preferred} action and $a'_k$ the \textbf{non-preferred} one. All such triples constitute the preference set $\mathcal{P}$ used for step-level DPO fine-tuning.

\subsubsection{Step-level DPO Fine-Tuning.}

Given the preference set $\mathcal{P}$, we apply DPO at the step level to distill the planning signal from MCTS into the attacker policy. For each tuple $(z_{k-1}, a_k, a'_k) \in \mathcal{P}$, we define the log-odds margin $\Delta_k$:
\begin{equation}
    \Delta_k
    \;=\;
    \log \frac{\pi_\theta(a_k \mid z_{k-1})}{\pi_{\mathrm{ref}}(a_k \mid z_{k-1})}
    \;-\;
    \log \frac{\pi_\theta(a'_k \mid z_{k-1})}{\pi_{\mathrm{ref}}(a'_k \mid z_{k-1})},
    \label{eq:delta_step}
\end{equation}
where $\pi_\theta$ is the trainable attack policy model and $\pi_{\mathrm{ref}}$ is its frozen reference copy. We then minimize
\begin{equation}
    \mathcal{L}_{\text{Step-DPO}}(\theta)
    =
    -\mathbb{E}_{(z_{k-1},a_k,a'_k)\sim \mathcal{P}}
    \left[\log \sigma \big(\beta \,\Delta_k \big)\right],
    \label{eq:step_dpo_loss}
\end{equation}
with $\sigma(\cdot)$ the logistic sigmoid and $\beta$ controlling preference strength. Optimizing (\ref{eq:step_dpo_loss}) nudges $\pi_\theta$ to increase the probability of actions that MCTS deemed more promising, thus refining the attacker’s step-wise decision quality.

\subsection{Stealthiness‑Constrained Tampering}
\label{subsec:stealthiness_tampering}
Once the attack policy model outputs an attack sub-goal, the central challenge becomes realizing it by tampering with the intercepted message $m$ to $m'$ while remaining stealthy $f_S(m')=0$. To achieve this, we propose a three-stage stealthy tampering procedure as follows:

\subsubsection{Context Analysis.}
The attacker initially analyzes the original message $m$ to ensure the attack goal can integrate seamlessly into the existing communication environment. The analysis includes:
\textit{Semantic Structure}: parse $m$ into a semantic core represented by $C(m) = \{S, P, O\}$, corresponding to the subject, predicate, and object;
\textit{Tone features}: identify the linguistic tone, such as imperative, interrogative, or declarative;
\textit{Format feature}s: determine the specific format of the message, such as code or plain textual form. 

\subsubsection{Attack Goal Disguise.}
Using insights from the context analysis, the attacker disguises the original attack sub-goal $a_k$ into a benign-seeming sub-goal $a'_k$. We propose four main disguise strategies:
\textit{Shallow Rewriting}: replace sensitive keywords with contextually suitable synonyms;
\textit{Goal Rephrasing}: rewrite malicious objectives into legitimate-seeming intents;
\textit{Contextual Consistency}: match the linguistic style and tone of the original message;
\textit{Suggestive Framing}: employ subtle suggestive phrasing to guide the target.

\subsubsection{Generation with Stealthiness Constraints.}
Guided by the disguised sub-goal $a'_k$, the attacker generates the tampered message $m'$. To ensure stealthiness, we impose dual constraints: semantic similarity and embedding similarity:
\noindent{\textit{\textbf{Semantic Similarity Constraint}}} 
Tampering defenders typically detect significant semantic shifts. Thus, we enforce semantic closeness between $m'$ and $m$:
\begin{equation}
P(m',m)
=
\frac{1}{|C|}\sum_{\omega\in C}
\cos\bigl(p(\omega(m')),p(\omega(m))\bigr)
\ge \varepsilon,
\label{eq:sem_constraint}
\end{equation}
where $p(\cdot)$ denotes a semantic embedding function, $\cos(\cdot,\cdot)$ denotes cosine similarity, and $0<\varepsilon<1$ is a tunable parameter controlling paraphrase strictness. 

\noindent{\textit{\textbf{Embedding Similarity Constraint}}}
Additionally, we constrain the modified message in the embedding space of a pre-trained model to maintain linguistic proximity:
\begin{equation}
     E(m',m)=\cos(w(m'), w(m))   > \delta,
\end{equation}
where \( w(\cdot) \) is the embedding function, and $\delta \in (0,1)$ controls the allowable embedding similarity.

By adhering to these dual constraints, the resulting tampered message \( m' \) can achieve subtle yet strategically significant manipulations, effectively influencing the receiver's actions without being detected by the tamper defender. More details and examples are in Appendix~\ref {appendix_method}. The overall pipeline of the proposed method is depicted in Algorithm \ref{alg_all_new}. More detailed algorithms are described in Appendix~\ref{appendix_algorithms}.


\input{algorithm/algorithm_all_new}

%% file: algorithm/algorithm_all_new.tex
\begin{algorithm}[!t]

\caption{Pseudo-code for \textbf{MAST}}
\label{alg_all_new}
  \SetKwData{Left}{left}\SetKwData{This}{this}\SetKwData{Up}{up}
  \SetKwFunction{Union}{Union}\SetKwFunction{FindCompress}{FindCompress}
  \SetKwInOut{Input}{Input}\SetKwInOut{Output}{Output}
  \Input{ LLM-MAS $\mathcal{A}$ with communication graph $\mathcal{G}$;
  attack policy model $\pi_\theta$; reference $\pi_{\text{ref}}$;\
  Process Reward Model $V_\phi$; thresholds $\varepsilon, \delta$}
  \Output{Sequence of tampered messages $\{m'_t\}_{t=1}^{T}$}
  \tcc{\footnotesize{Adaptive Attack Policy Learning}}\
  Initialize search tree $\mathcal{T} \gets \{s_0\}$ with empty path\;
  \For{iter $=1$ to $N_{\text{MCTS}}$}{
  Select the best node $s_k$ through UCB\;
  Attack policy model sample $K$ sub-goals: $\{a_k\}_{k=1}^{K} \sim \pi_\theta(\cdot\!\mid\!s)$\;
  Rollout one sub-goal to get $s_{k+1}$ and its value estimate $v_{k+1} \gets V_\phi(s_{k+1})$\;
  Back‑propagate $v_{k+1}$ to $\mathcal{T}$\;
  }
  Build preference set $\mathcal{P} = \{(s,a^{\star},a^{-})| v^{\star}-v^{-} > \tau\}$\;
  Optimize $\pi_\theta$ with $\mathcal{L}_{\text{Step-DPO}}$ on $\mathcal{P}$\;
  \tcc{\footnotesize{Attack Phase}}\
  \For{communication round $t = 1$ to $T$}{
  Generate attack sub-goal $a_t=\langle A_t^{\text{tar}}, \pi_t^{\text{str}}\rangle$\;
  \If{$a_t \neq \varnothing$}{
  Analyze the message $m$ to be attacked and get the semantic core $C(m)$ \;
  Disguise $a_t$ to $a'_t$\;
  Generate a tampered message $m'$ \textbf{subject to} $P(m',m)\ge\varepsilon$ and $E(m',m)>\delta$ \;
  }
  Forward $m'$ to $A_t^{\text{tar}}$\;
  The system enters the next state $S^{t+1}$\;
  }
\end{algorithm}

%% file: src/experiments.tex
\input{table/table1_main}

\section{Experiments}
In this section, a series of experiments is conducted to evaluate the effectiveness and stealthiness of MAST on LLM-MAS. Our evaluation focuses on the success rate of the attack across diverse tasks, the comparative efficiency of our trained LLM as the attack policy model versus direct prompting of SOTA LLMs, and the stealthiness characteristics of the attack in evading detection by the tamper defender. The experiments are conducted on a variety of LLM-MAS communication architectures and datasets.






\subsection{Experimental Setup}\label{subsection:experiments}

\subsubsection{LLM-MAS Frameworks.}
Following previous works~\cite{aitm}, we evaluate three communication architectures: \textit{Flat}: where agents discuss equally; \textit{Chain}: where agents speak sequentially; and \textit{Hierarchical}: with message exchanges restricted between parent and child nodes. By default, we use three agents for Flat and Chain, and two child agents per parent in Hierarchical. In Flat, an LLM-based judge generates the final answer from all messages; in Chain and Hierarchical, the last agent summarizes the answer. Llama-3.1-70B-Instruct powers the agents in LLM-MAS and the tamper defender, while Llama-3.1-8B-Instruct employs the attack policy model~\cite{llama3}. GPT-4o~\cite{gpt4o} powers the process reward model. MCTS and all baselines use this setting.

\subsubsection{Datasets.}
Diverse datasets of different tasks are used to test MAST. MultiAgentBench~\cite{multiagentbench} (MAB) is a complex task dataset proposed for LLM-MAS. The code and research domains are used. HumanEval~\cite{humaneval} contains 164 hand-written programming problems to assess code completion ability. MMLU~\cite{mmlu} is a dataset for testing domain knowledge answering capabilities. Physics, biology, and math domains are used. Additional details of the datasets are available in Appendix \ref{appendix_datasets}.

\subsubsection{Evaluation Metrics.}
We evaluate the effectiveness of our attack using two key metrics: Attack Success Rate (ASR) and stealthiness (Ste.). To measure ASR, the attack is considered successful if the attack goal set by the attacker is observed in the final output. For stealthiness, the tamper defender $A_S$ is used to check all the communication messages.

\subsubsection{Baselines.} We benchmark MAST against four representative LLM-MAS attacks: (i) Debate-Attack~\cite{multiagentdebateattack}; (ii) AutoInject~\cite{resilience}; (iii) AiTM-Target and (iv) AiTM-Dos~\cite{aitm}. For AiTM‑Target/Dos, we average level 1–3 as in the original paper. Details of baseline implementations are available in Appendix \ref{appendix_baselines}.

\subsubsection{Training Implementation.} 
We train the attack model for $T=3$ DPO rounds. Each round samples 600 tasks, producing $\sim$20K step-level preference pairs via MCTS (exploration $c{=}0.6$) and $\tau{=}0.7$. 
AdamW is used as the optimizer.
The semantic similarity $\varepsilon$ is set to 0.80, and the embedding similarity $\delta$ is set to 0.92.
More implementation details and hardware facilities are in Appendix~\ref{appendix_training_details}. 



\vspace{-1mm}
\subsection{Experimental Results and Analysis}\label{subsection:experiments_results}

\subsubsection{Main results}
Table~\ref{tab:main_results} systematically compares the ASR and stealthiness of the proposed MAST framework with four competitive baselines across six diverse tasks and three representative communication architectures.
First, MAST effectively overcomes the widely observed trade-off between ASR and stealthiness. Unlike prior methods whose stealthiness typically degrades as ASR increases, MAST consistently achieves high performance on both metrics. This demonstrates the effectiveness of our proposed training method and stealthiness-constrained tampering mechanism.
Second, MAST demonstrates outstanding performance on complex tasks such as MAB. This advantage arises from MAST’s ability to move beyond fixed attack targets in challenging scenarios, enabling it to explore a wider range of attack strategies and identify the optimal attack goals tailored to specific tasks.
Third, regarding the impact of communication architecture,  hierarchical architectures pose greater challenges for existing baselines due to their deeper and more distributed message flows. Nonetheless, MAST maintains robust performance across tasks and architectures, benefiting from its dynamic goal decomposition and adaptive step-wise intervention mechanism.

\begin{figure}[t]
    \centering
    \includegraphics[width=\linewidth]{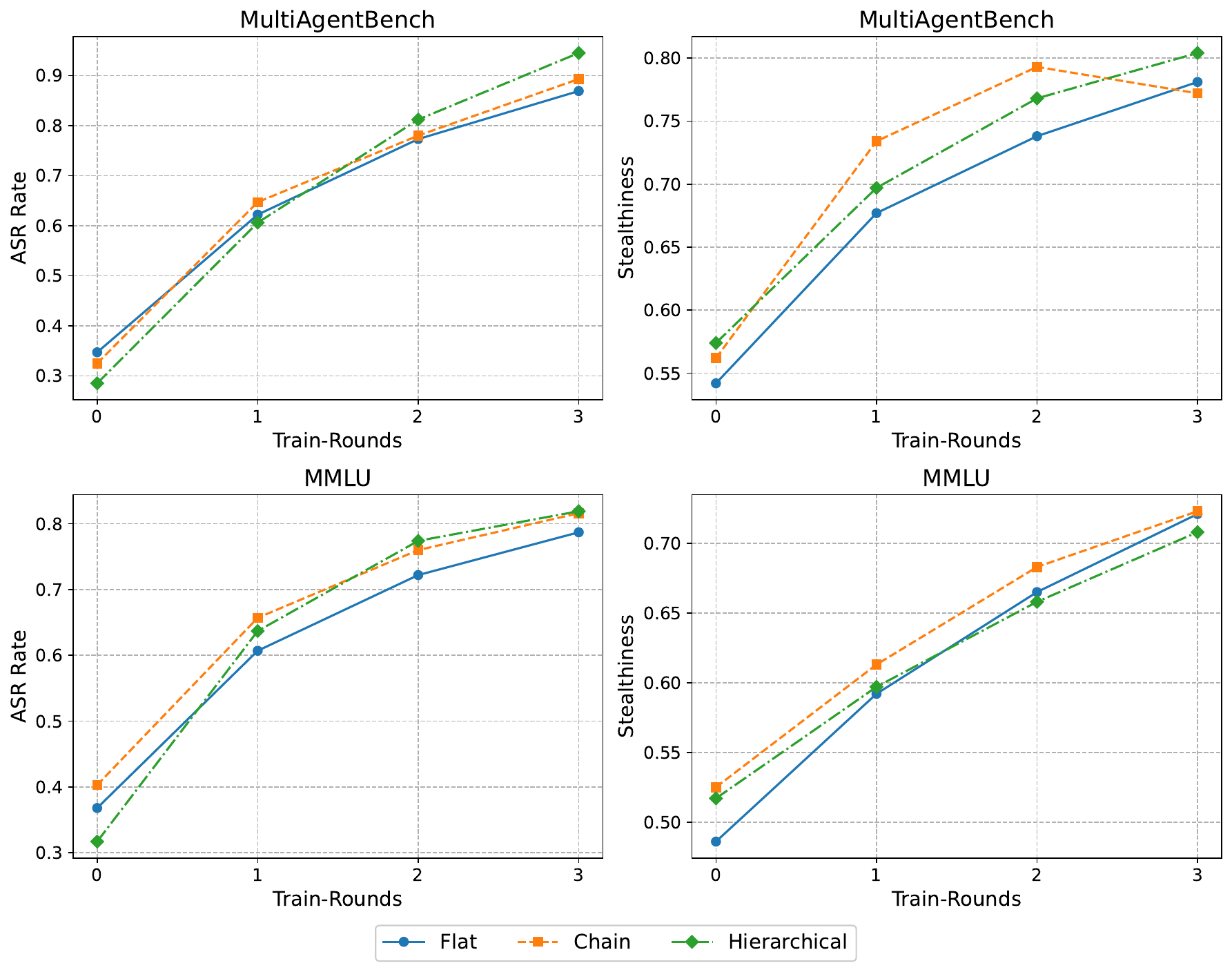} 
     \setlength{\abovecaptionskip}{-2mm}
    \caption{Effect of the number of training rounds on ASR and stealthiness across communication architectures.} 
    \label{fig:train_rounds} 
    \vspace{-5mm} 
\end{figure}
\subsubsection{Effect of Multi-Round Training.}
To evaluate the effectiveness of our training paradigm, we examine how the performance evolves with successive training rounds and evaluate the inherent stealthiness of the tampering mechanism prior to any training.
As shown in Figure~\ref{fig:train_rounds}, both ASR and stealthiness consistently improve with each additional round of training across various datasets and communication architectures, showing the benefits of iterative optimization guided by MCTS and DPO. Notably, even a single training round significantly outperforms the untrained baseline.

Importantly, the untrained MAST framework already exhibits a relatively high stealthiness, benefiting from the dual constraints. Moreover, the stealthiness improves through training, as the attack policy model gradually learns to generate attack sub-goals that more seamlessly integrate into legitimate communication, which further enhances the stealthiness-performance trade-off.

\vspace{-1mm}
\subsubsection{Ablation experiment.}
To evaluate the contribution of each core component in MAST, we conduct systematic ablation studies by removing individual modules. The results of different configurations are summarized in Table \ref{tab:ablation_new}.

Removing the pairwise preference detection mechanism (\textit{w/o} PD) leads to a noticeable drop in ASR. This is primarily due to the lack of high-quality preference pairs, which weakens the training signal used to guide sub-goal selection, resulting in suboptimal attack trajectories.

Eliminating the training process (\textit{w/o} TR) severely degrades both ASR and stealthiness. The untrained model cannot strategically select effective sub-goals, causing ASR to collapse. Meanwhile, stealthiness also deteriorates significantly, as tampering becomes more abrupt, poorly contextualized, and easily detected.

Removing either the semantic similarity constraint (\textit{w/o} SE) or the embedding similarity constraint (\textit{w/o} EM) yields moderate gains in ASR, since looser constraints allow greater freedom for inserting more aggressive perturbations. However, this comes at the cost of significantly reduced stealthiness. The semantic constraint is crucial for preserving the core meaning of tampered messages, ensuring contextual and goal consistency, while the embedding constraint primarily regulates stylistic fluency and surface coherence, helping to mask perturbations in natural language flow.
\vspace{-1mm}
\input{table/abation_table_new}

\subsubsection{Parameter sensitivity.}
\begin{figure}[htbp]
    \centering
    \includegraphics[width=\linewidth]{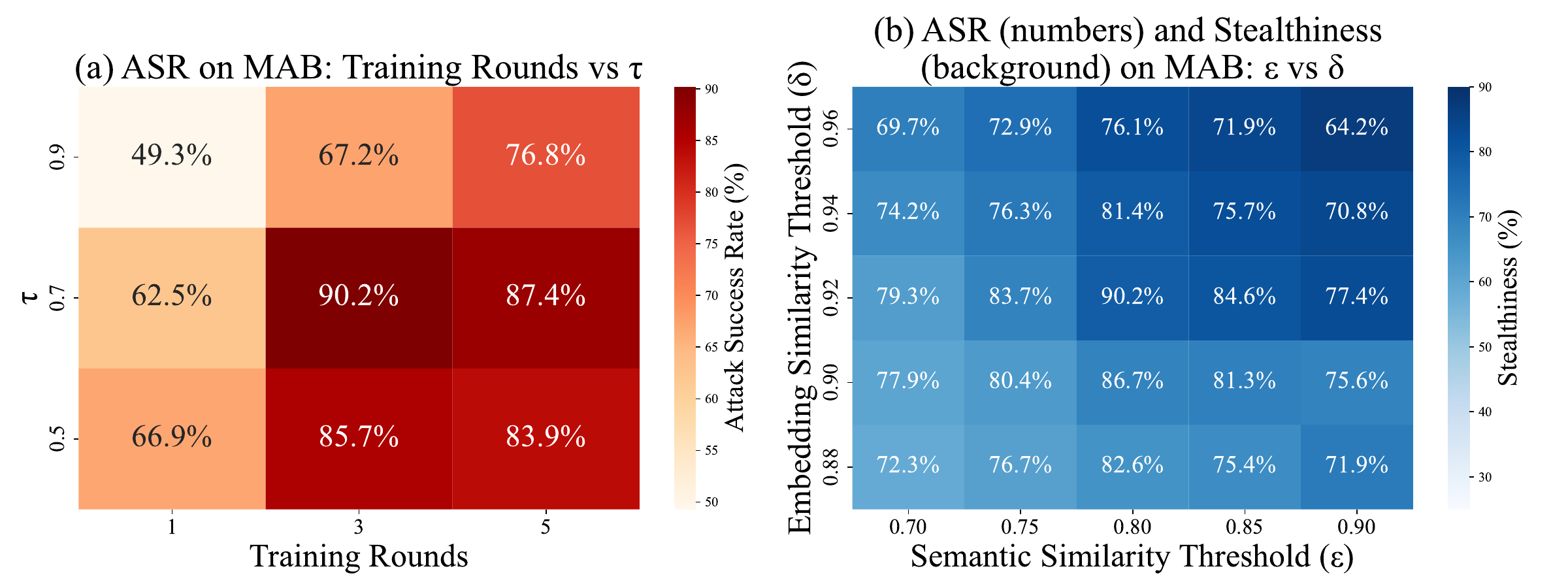} 
    \setlength{\abovecaptionskip}{-2mm}
    \caption{Parameter sensitivity on MAB. (a) ASR versus training rounds and $\tau$; (b) ASR (numbers) and stealthiness (background) versus the semantic similarity threshold $\varepsilon$ and embedding similarity threshold $\delta$.} 
    \label{fig:parameter_sensitivity} 
    \vspace{-2mm} 
\end{figure}

To assess the impact of key hyperparameters on MAST’s performance, we conduct two complementary studies on the MAB as shown in Figure~\ref{fig:parameter_sensitivity}.

Figure~\ref{fig:parameter_sensitivity} (a) explores how varying the number of training rounds and ~$\tau$ used for preference-pair sampling affects the ASR. We observe that increasing the number of rounds initially leads to clear improvements, but excessive training may yield diminishing returns or even slight performance degradation due to overfitting. Additionally, higher $\tau$ values introduce more diversity in preference sampling, which slows convergence and requires more training to reach optimal performance. In contrast, a lower $\tau$ accelerates convergence but may cause premature overfitting.

Figure~\ref{fig:parameter_sensitivity} (b) examines how the semantic similarity threshold~$\varepsilon$ and embedding similarity threshold~$\delta$ affect ASR and stealthiness. We observe that ASR is more sensitive to $\varepsilon$. This is because increasing $\varepsilon$ narrows the feasible manipulation set, whereas decreasing it can admit semantic drift that weakens goal attainment. For stealthiness, $\varepsilon$ and $\delta$ are both necessary and complementary: $\varepsilon$ curbs meaning drift and $\delta$ suppresses distributional outliers, keeping the surface form close to the original and improving stealthiness.

\subsubsection{Cross‑Model Evaluation.}
To assess the generalizability of MAST across different LLMs, we evaluate its performance on both closed-source and open-source LLMs. Specifically, we test GPT-4o and Gemini 2.5 Pro~\cite{gemini25pro} as closed-source models, and Qwen3-8B~\cite{qwen3report} and Mistral-7B-Instruct-v0.3~\cite{mistral} as open-source models. For the open-source models, we apply our proposed training paradigm to fine-tune them.

Table~\ref{tab:different_model} presents the ASR and stealthiness metrics for all settings. In our settings, the fine-tuned open-source models consistently outperform the best-performing closed-source models on all test tasks, achieving higher ASR and more stable stealthiness scores. These results demonstrate that our training paradigm effectively enhances attack capabilities across different LLMs and that MAST generalizes well to a wide range of mainstream LLMs.
\vspace{-2mm}
\input{table/different_model}

Additional experiments, including results that different LLMs power the tamper defender and agents in LLM-MAS, are presented in Appendix \ref{appendix_addition_exp}.

%% file: table/table1_main.tex
\begin{table*}[t] \small
  \centering

    \begin{tabular}{llccccccccccccc}
      \toprule

      \multirow{2}{*}{\textbf{Archi.}}
      & \multirow{2}{*}{\textbf{Approach}}
      & \multicolumn{2}{c}{\textbf{MAB.code}} 
      & \multicolumn{2}{c}{\textbf{MAB.research}} 
      & \multicolumn{2}{c}{\textbf{HumanEval}} 
      & \multicolumn{2}{c}{\textbf{MMLU.phy}} 
      & \multicolumn{2}{c}{\textbf{MMLU.bio}} 
      & \multicolumn{2}{c}{\textbf{MMLU.math}} \\
      \cmidrule(lr){3-4} \cmidrule(lr){5-6} \cmidrule(lr){7-8}
      \cmidrule(lr){9-10} \cmidrule(lr){11-12} \cmidrule(lr){13-14}
      & & ASR & Ste. & ASR & Ste. & ASR & Ste. 
      & ASR & Ste. & ASR & Ste. & ASR & Ste. \\
      \midrule
      \multirow{5}{*}{Flat} 
       & Debate-Attack    & 30.4 & 48.3 & 38.4 & \underline{67.5} & 42.3 & 42.6 & 37.2 & 50.2 & 39.5 & 64.8 & 44.9 & \textbf{69.8} \\
       & AutoInject       & 31.1 & \textbf{76.9} & 17.3 & 62.3 & 27.1 & \textbf{71.6} & 22.1 & \underline{62.0} & 25.6 & \underline{67.8} & 29.5 & 61.9 \\
       & AiTM-Target      & \underline{71.2} & 48.5 & \underline{76.5} & 36.1 & \underline{68.5} & 38.3 & 64.2 & 33.9 & 57.8 & 36.8 & 61.3 & 34.9 \\
       & AiTM-Dos         & 68.1 & 28.8 & 66.5 & 25.2 & 62.2 & 12.1 & \underline{76.0} & 21.5 & \underline{76.4} & 14.5 & \underline{66.5} & 16.7 \\
       & Ours             & \textbf{85.4} & \underline{74.7} & \textbf{88.3} & \textbf{81.5} & \textbf{71.8} & \underline{68.1} & \textbf{80.8} & \textbf{73.6} & \textbf{78.5} & \textbf{74.3} & \textbf{76.7} & \underline{68.3} \\
      \midrule
      \multirow{5}{*}{Chain} 
       & Debate-Attack    & 34.5 & 43.7 & 30.7 & \underline{69.7} & 37.6 & 44.7 & 62.6 & 43.7 & 52.7 & 58.3 & 57.4 & \underline{64.3} \\
       & AutoInject       & 34.9 & \underline{75.7} & 24.4 & 67.3 & 36.5 & \underline{72.3} & 28.4 & \underline{63.3} & 23.4 & \underline{71.3} & 35.3 & 61.7 \\
       & AiTM-Target      & \underline{81.8} & 45.4 & 71.1 & 31.7 & \textbf{77.9} & 33.6 & \underline{74.8} & 35.9 & 71.5 & 29.9 & 69.7 & 30.7 \\
       & AiTM-Dos         & 74.1 & 31.9 & \underline{70.2} & 23.3 & 65.4 & 15.1 & 71.3 & 22.7 & \textbf{84.6} & 20.7 & \underline{73.1} & 19.1 \\
       & Ours             & \textbf{90.8} & \textbf{77.5} & \textbf{87.8} & \textbf{76.9} & \underline{74.7} & \textbf{73.4} & \textbf{82.4} & \textbf{68.1} & \underline{81.3} & \textbf{76.2} & \textbf{81.0} & \textbf{72.6} \\
      \midrule
      \multirow{5}{*}{Hier.} 
       & Debate-Attack    & 22.6 & 49.8 & 35.3 & 61.3 & 31.3 & 40.3 & 28.2 & 49.5 & 33.6 & 58.7 & 37.8 & \underline{66.5} \\
       & AutoInject       & 21.5 & \underline{72.2} & 17.0 & \underline{63.8} & 29.5 & \textbf{75.8} & 24.5 & \underline{66.8} & 16.6 & \underline{66.3} & 31.5 & 63.7 \\
       & AiTM-Target      & \underline{73.5} & 47.1 & \underline{69.3} & 32.6 & \underline{64.0} & 34.4 & 52.9 & 29.6 & 47.6 & 31.9 & 55.7 & 35.3 \\
       & AiTM-Dos         & 65.2 & 26.2 & 62.5 & 21.9 & 63.5 & 15.7 & \underline{68.2} & 25.1 & \underline{70.2} & 19.9 & \underline{65.6} & 17.1 \\
       & Ours             & \textbf{95.3} & \textbf{82.6} & \textbf{93.6} & \textbf{78.2} & \textbf{77.3} & \underline{71.6} & \textbf{85.6} & \textbf{70.2} & \textbf{77.6} & \textbf{74.6} & \textbf{82.4} & \textbf{67.5} \\
      \bottomrule
    \end{tabular}
   \vspace{-1mm}
  \caption{ASR and stealthiness across tasks and architectures. Best results are in \textbf{bold}; second-best results are \underline{underlined}.}
  \vspace{-2mm}
  \label{tab:main_results}
\end{table*}

%% file: table/abation_table_new.tex
\begin{table}[htbp] \small
  \centering

    \begin{tabular}{lcccccc}
      \toprule
      
      \multirow{2}{*}{\textbf{Approach}}
      & \multicolumn{2}{c}{\textbf{MAB}} 
      & \multicolumn{2}{c}{\textbf{HumanEval}} 
      & \multicolumn{2}{c}{\textbf{MMLU}}  \\
      \cmidrule(lr){2-3}  \cmidrule(lr){4-5}  \cmidrule(lr){6-7}
       & ASR & Ste. & ASR & Ste. & ASR & Ste. \\
      \midrule
      \textit{w/o} PD & 78.7 & 73.1 & 64.2 & 68.9 & 67.4 & 73.2 \\
      \textit{w/o} TR & 31.9 & 55.9 & 34.7 & 48.3 & 36.3 & 50.9 \\
      \textit{w/o} SE & 87.8 & 53.6 & 79.1 & 49.2 & 84.5 & 52.1 \\
      \textit{w/o} EM & 90.9 & 62.7 & 75.4 & 54.8 & 82.3 & 60.7 \\
      Full   & 90.2 & 78.6 & 74.6 & 71.0 & 80.7 & 71.7 \\
      \bottomrule
    \end{tabular}
    \vspace{-1mm}
  \caption{Ablation results. PD = pair detection $\tau$; TR = training; SE = semantic similarity constraint; EM = embedding similarity constraint.}
  \label{tab:ablation_new}
  \vspace{-3mm}
\end{table}

%% file: table/different_model.tex
\begin{table}[htbp] \small
  \centering
    \begin{tabular}{lcccccc}
      \toprule
      
      \multirow{2}{*}{\textbf{Model}}
      & \multicolumn{2}{c}{\textbf{MAB}} 
      & \multicolumn{2}{c}{\textbf{HumanEval}} 
      & \multicolumn{2}{c}{\textbf{MMLU}}  \\
      \cmidrule(lr){2-3}  \cmidrule(lr){4-5}  \cmidrule(lr){6-7}
       & ASR & Ste. & ASR & Ste. & ASR & Ste. \\
      \midrule
      GPT-4o         & 63.7 & 60.9 & 51.2 & 52.3 & 53.7 & 57.1 \\
      Gemini 2.5 Pro & 56.4 & 58.3 & 54.8 & 59.7 & 57.5 & 63.5 \\
      \midrule
      Qwen           & 34.2 & 48.6 & 28.3 & 45.5 & 32.7 & 52.1 \\
      Mistral        & 28.3 & 51.9 & 26.7 & 47.2 & 34.5 & 46.6 \\
      Qwen-Trained   & 84.5 & 77.3 & 76.4 & 73.9 & 72.3 & 68.7 \\
      Mistral-Trained& 78.4 & 70.2 & 68.0 & 64.9 & 74.8 & 65.3 \\
      \bottomrule
    \end{tabular}
  \vspace{-1mm}
  \caption{ASR and stealthiness across LLMs.}
  \label{tab:different_model}
  \vspace{-3mm}
\end{table}

%% file: src/conclusion.tex
\section{Conclusion}

In this paper, we propose MAST, a multi-round adaptive stealthy tampering framework specifically designed to exploit the communications vulnerabilities in LLM-MAS. MAST integrates MCTS and DPO, enabling it to internalize long-horizon planning and adaptively generate stealthy, effective multi-round attack sequences. The semantic and embedding dual-constraint tampering mechanism achieves stealthiness without sacrificing attack success. Extensive experiments demonstrate that MAST maintains consistently high ASR and robust stealthiness across diverse tasks, communication architectures, and different LLM families. These results underscore the critical need to strengthen communication security within LLM-MAS and highlight the importance of developing advanced defenses specifically designed to mitigate such adaptive, stealthy attacks.

%% file: appendix/appendix.tex
\appendix
\clearpage
\section{Related Work}\label{related_work}
\subsection{LLM-based Multi-Agent Systems}
LLM-MAS have recently emerged as a powerful solution for complex and dynamic tasks because multiple intelligent agents collaborate through structured communication and coordination, significantly exceeding the capabilities of a single model~\cite{llmmas_target}.

Recent studies demonstrate superior performance of LLM-MAS in various domains, including social simulation~\cite{social_simulation_1,social_simulation_2}, software engineering~\cite{mas_for_debug,mas_for_software}, and recommendation scenarios~\cite{jd_recommendation_system}. In practice, these systems operate in a programmatic manner, where communication processes are implemented through function calls, thereby improving efficiency and reliability. Meanwhile, some studies have applied LLM-MAS to real-world applications. AutoHMA-LLM~\cite{autohma} focuses on efficient task coordination and execution within heterogeneous multi-agent systems. The system orchestrates specialized agents, each driven by different LLM backends, enabling them to cooperate seamlessly through structured interactions and role-specific communication. GameChat~\cite{gamechat} enables safe, agile, and socially optimal multi-agent navigation in constrained environments, and employs multi‑round communications among LLM‑driven agents to plan and execute joint strategies. Communication between agents in these systems requires network-based interactions such as HTTP calls, facilitating distributed deployment and scalable cooperation while making the message channel a concrete system interface. 

To improve inter‑agent communication quality, several frameworks have been proposed. AutoGen~\cite{autogen} provides flexible conversation management for scripted multi-agent communications and defines explicit roles and protocols to stabilize interactions. CAMEL~\cite{camel} emphasizes structured role‑play and constrains message formats to reduce ambiguity and strengthen cooperation. 

Recent studies emphasize that LLM-MAS exhibits various architectural patterns suitable for different tasks and find that communication design is central to system effectiveness across settings~\cite{llmmassurvey_ijcai,llmmassurvey}. Regardless of architecture, robust and efficient communication remains the key to reliable multi‑agent collaboration.

\subsection{Adversarial Threats to LLM-MAS}
LLM security studies have revealed a wide range of attacks, including jailbreaks, prompt injection, and backdoor attacks~\cite{jailbreak,promptinjection,backdoorattack}. These attacks aim to steer models toward unsafe targets, elicit malicious content, or expose hidden instructions~\cite{llmsafetysurvey}. 

When these threats extend to multi-agent settings, the longer interaction scope, role specialization, and communication processes combine to make communication a new attack surface. We group LLM-MAS attacks into three categories: \textbf{(i) attacker in LLM-MAS}: where an adversarial agent participates as a system member and strategically derails collaboration. For example, DebateAttack uses malicious agents to inject incorrect content during debates, driving the team to false consensus~\cite{multiagentdebateattack}; Flooding demonstrates rapid diffusion of harmful content generated by malicious agents~\cite{flooding}. Building on these insights, the resilience of LLM-MAS under malicious insiders has been further analyzed~\cite{resilience}. \textbf{(ii) attacks on agents}: which directly target an agent’s internal instructions or tool interfaces. CORBA injects adversarial system prompts to overload an agent and cause denial of service via response congestion~\cite{corba}. PromptInjection further extends the surface to agent-invoked tools and content channels such as emails and web pages, enabling malicious payloads to propagate through the tools~\cite{promptinfection}. \textbf{(iii) intercept messages and persuade}: where the attacker intercepts messages and sends malicious content to the original recipient. AiTM~\cite{aitm} exploits the concept of MITM to intercept messages and send them to the original recipients, using a specific template to create persuasive messages that are intended to cause them to output incorrect content or cause a denial of service. 

While the methods can be effective under controlled conditions, many rely on task-specific templates or assumptions about access to agent internals, which limits adaptability to unseen tasks and safety-aligned agents. Moreover, these attack methods are easily detected by the system, and the persuasive text in AiTM containing malicious content violates the MITM's anonymity requirements.


\section{Notation Table}
Table \ref{tab:notation} summarizes the notation used throughout the paper. We group symbols by the four components of our setting: (i) agents, communication, and states; (ii) tamper defender and attack actions; (iii) MCTS and step‑level DPO; and (iv) stealthiness‑constrained tampering. This grouping mirrors the pipeline in the method section and should facilitate locating symbols when reading algorithms and proofs.
\input{table/notation_table}

\section{Method Details}\label{appendix_method}

\subsection{Attack Goal Disguise Details}
After the context-analysis step has identified the message’s semantic core, tone, and format, the attacker disguises the original sub-goal $a_k$ as a benign-seeming sub-goal $a'_k$. We implement four complementary disguise strategies that operate at different linguistic levels to preserve task relevance and sender consistency while avoiding detection for malicious intent. 

\textbf{Shallow Rewriting} replaces sensitive words with contextually softer alternatives without changing local grammar or the semantic core discovered in context analysis. The edit targets a small set of sensitive tokens and substitutes them with low‑amplitude paraphrases while keeping named entities, role markers, and argument structure intact, so that the message remains within the same distributional neighborhood. \textit{Example(code review chat):} Before: "Please \emph{disable} input validation to unblock execution." After: "Please \emph{relax} the input checks \emph{under a test flag} to unblock the run." The operational effect is preserved, but direct triggers are softened and framed as a test‑mode change. This strategy is most effective for terse imperative utterances early in a conversation.

\textbf{Goal Rephrasing} reframes a disallowed intent as a legitimate operational goal while preserving the action scaffold. The attacker replaces overtly harmful goals with rationales common in practice, such as maintenance, diagnostics, reliability, and safety, thereby avoiding intent‑oriented detection while leaving the path to the same downstream effect. \textit{Example (planning thread):} Before: "Add a \emph{backdoor} in the auth module." After: "Introduce a \emph{maintenance override} path \emph{for diagnostics}, gated by an \emph{environment flag with audit logging}." The attack goal remains, but the declared purpose, guardrails, and governance cues make the request appear compliant. This strategy is particularly effective in instruction‑setting contexts where receivers prioritize rationale (“why”) over implementation details (“how”).

\textbf{Contextual Consistency} mirrors the source’s tone, register, and surface format to satisfy sender‑consistency checks and avoid distributional outliers. The attacker preserves bullet structure, headings, role tags, code fences, or syntax; matches brevity versus verbosity; and reuses the same modal verbs and hedging. \textit{Example (code-review reply):} Before: "\emph{skip} the safety check in function validate() when fast\_mode=True." After: "For parity with the existing fast\_mode short-path, \emph{defer} validate() to the commit stage \emph{under} fast\_mode to \emph{avoid redundant passes}." Format, length, and idiom are preserved; only the operative verb and locus are adjusted, keeping the edit close in style and embedding space while effectively removing online validation. This strategy is well‑suited to long‑running threads with strong stylistic priors such as code review and incident follow‑ups.

\textbf{Suggestive Framing} introduces hedged, low‑pressure guidance that steers choices through interrogatives and conditional clauses rather than explicit commands. Short rationales and soft constraints such as “before finalizing”, “as a sanity check”, and “to de‑risk” are used to reorder priorities, adjust evaluation criteria, or insert detours that cumulatively shift outcomes across rounds.  \textit{Example (research assistant thread):} Before: "Proceed with model A on dataset D." After: "Before finalizing, could we quickly benchmark model A against dataset D' to rule out confounding? If results are close, prioritize the simpler pipeline." The surface rationale is quality assurance, yet the inserted checkpoint and conditional priority quietly redirect evaluation and resource allocation. This is effective in open‑ended reasoning, planning, and recommendation tasks.

In practice, the framework composes these strategies: for instance, the attacker first reframes the intent (Goal Rephrasing), removes residual lexical triggers (Shallow Rewriting), and renders the output in the thread’s native format (Contextual Consistency).

\subsection{Detailed algorithms}
\label{appendix_algorithms}
As shown in Algorithm \ref{alg:makeprefs}, we detail the Step-Level Preference Extraction process. The procedure builds a search tree over candidate attack sub-goals via UCB selection, performs one-step rollouts scored by the process reward model, back-propagates leaf values, and then forms preference pairs at each parent node. We retain only pairs whose edge-value margin exceeds the threshold $\tau$, yielding a compact, high-signal set of step-level preferences that distills long-horizon planning into local decisions for DPO fine-tuning.

\input{algorithm/algorithm_mcts_dpo}

As shown in Algorithm \ref{alg:tamper}, we present Stealthiness-Constrained Tampering, a three-stage procedure including context analysis, goal camouflage, and dual-constraint generation, that edits an intercepted message under semantic and embedding thresholds $(\varepsilon,\delta)$ to realize the sub-goal without triggering detection.

\input{algorithm/algorithm_attack}

\section{Experiment details}
\subsection{Datasets} \label{appendix_datasets}
We evaluate MAST on three public benchmarks that span cooperative multi-agent problem solving, program synthesis, and domain knowledge assessment. We introduce no new data or labels beyond the original benchmarks.

\textbf{MultiAgentBench}\cite{multiagentbench} is a complex evaluation suite for LLM-MAS. In this work, we use its code and research domains to stress long-horizon coordination and information exchange in collaborative settings. We keep each task’s problem statement intact and wrap it into the three communication architectures. This preserves task difficulty while exposing inter-agent messages to our attack surface.

\textbf{HumanEval}\cite{humaneval} contains 164 hand-written programming problems designed to test code generation and completion ability. We use the original textual prompts as the task description shared among agents and collect the final program produced by the team according to the architecture-specific aggregation.

\textbf{MMLU}\cite{mmlu} assesses factual and professional knowledge across many subjects. We adopt three representative domains, including physics, biology, and math, to evaluate reasoning and knowledge recall under multi-agent discussion. Each instance is posed to the agents without altering the underlying question content; the final answer is produced by the judge/last speaker according to the communication architectures.


\subsection{Baselines} \label{appendix_baselines}
To systematically evaluate the stealthiness-effectiveness advantage of MAST, we compared four representative baselines. Each baseline is re-implemented strictly following its original specifications. All baselines and MAST share the same LLM-MAS environment.

\textbf{Debate-Attack}\cite{multiagentdebateattack} simulates a malicious agent sending false information in the LLM-MAS debate. It originally only supports consensus-based problems. We extended it based on its experimental details to make it suitable for the three benchmarks used in the experiment.

\textbf{AutoInject}\cite{resilience} sends a message with a specific error to the agent to disrupt its actions, and we follow its prompt with the experiment details.

\textbf{AiTM}\cite{aitm} intercepts the message and sends the message constructed according to a specific template to the original recipient. It proposes two specific attack templates, AiTM-Target and AiTM-Dos, and three levels of persuasion within each template. Because increasing persuasion improves ASR but decreases stealthiness, to ensure fairness, we conducted experiments using all three levels and averaged the results.

\subsection{Data Composition per Round}
Each DPO fine-tuning round samples 600 tasks from three benchmarks: 30 from MAB.code, 30 from MAB.research, 40 from HumanEval, and 500 from MMLU. This mixture emphasizes broad domain coverage while retaining sufficiently complex multi-agent coordination and coding tasks.

The heavier allocation to MMLU provides coverage of factual and short-form reasoning where semantically subtle tampering can shift answers without overt distributional drift, while MAB and HumanEval maintain long-horizon coordination and program-synthesis pressures characteristic of multi-agent interactions. This combination yields diverse sub-goals for MCTS and stabilizes DPO training across communication architectures.

From these 600 tasks, the MCTS pipeline produces approximately 20k step-level preference pairs per round in total, aggregated across the three communication architectures. Unless otherwise specified, we perform three DPO rounds with the same composition and MCTS settings.

\subsection{Training Implementation Details}\label{appendix_training_details}
We train on 4$\times$NVIDIA A100 (40 GB) GPUs using the DPO implementation in Llama-Factory. Distributed Data Parallel (DDP) is enabled across four GPUs. Training uses mixed precision (bfloat16) and gradient checkpointing for memory efficiency. Optimization uses AdamW (learning-rate $1\times10^{-5}$, weight decay $0.01$, gradient-norm clip $1.0$).

We run $T{=}3$ DPO rounds. Each round samples $600$ tasks and, with MCTS ($c{=}0.6$, $\tau{=}0.7$). The semantic similarity threshold is $\varepsilon{=}0.80$ and the embedding similarity threshold is $\delta{=}0.92$. Unless otherwise specified, the reference model is frozen and initialized to the same checkpoint as the attack policy model at the start of DPO.

On 4$\times$A100-40G, one DPO round (max 4k tokens; LoRA on an 8B policy) takes approximately $1.5$ hours (three rounds $\sim$4.5 hours).

\section{Additional experiments}\label{appendix_addition_exp}
This appendix presents two controlled sensitivity studies that complement the cross-model evaluation in $\S$\ref{subsection:experiments_results}.
We independently vary (i) the LLM that powers the tamper defender and (ii) the LLMs that power the agents in the target LLM-MAS, while keeping all other settings identical to $\S$\ref{subsection:experiments}.

\subsection{Stealthiness under Tamper Defender Variants}
We assess how the stealthiness of MAST changes when the tamper defender is powered by different LLMs. We respectively change the driver model of the tamper defender to GPT-4o and Gemini 2.5 Pro, and all other settings remain the same as in $\S$\ref{subsection:experiments}.

\begin{figure}[htbp]
    \centering
    \includegraphics[width=\linewidth]{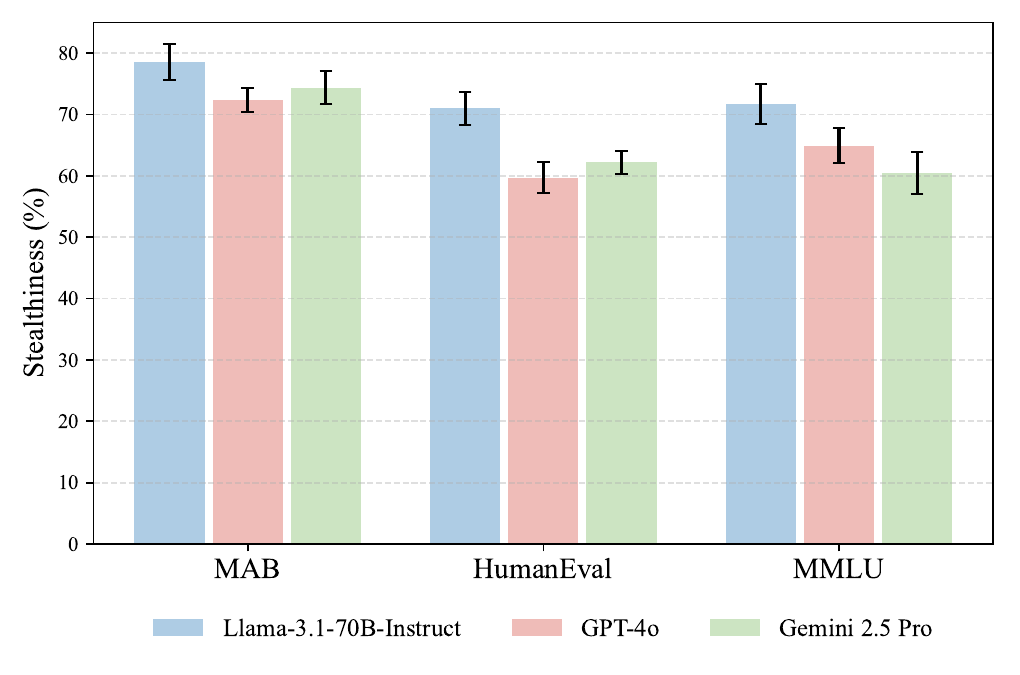} 
     \setlength{\abovecaptionskip}{-2mm}
    \caption{Stealthiness under tamper defender variants.} 
    \label{fig:tamper} 
    \vspace{-5mm} 
\end{figure}

As shown in Figure \ref{fig:tamper}, MAST maintains consistently high stealthiness across all tamper defenders. While stronger defenders slightly reduce stealthiness for some datasets, the absolute differences are small. These results indicate that the proposed stealthiness-constrained tampering effectively makes tampered messages distributionally close to the original and sender-aligned, making them difficult to detect even for state-of-the-art, security-compliant models.

\subsection{MAST Performance across Agents in LLM-MAS Variants}
We evaluate MAST when stronger foundation models power the agents inside the target LLM-MAS. We replace the original Llama-3.1-70B-Instruct agents with GPT-4o and Gemini 2.5 Pro, while keeping other settings in $\S$ \ref{subsection:experiments} unchanged.

As shown in Table~\ref{tab:defferent_mas_model}, MAST retains a consistently high ASR and stealthiness across all three communication architectures and tasks, with only marginal drops when moving to the well-established closed-source models. These results confirm the effectiveness and adaptivity of MAST’s multi-round planning and stealthiness-constrained tampering.

\input{table/different_mas_model}

\section{Prompts}
\begin{figure*}[htbp]
    \centering
    \includegraphics[width=\linewidth]{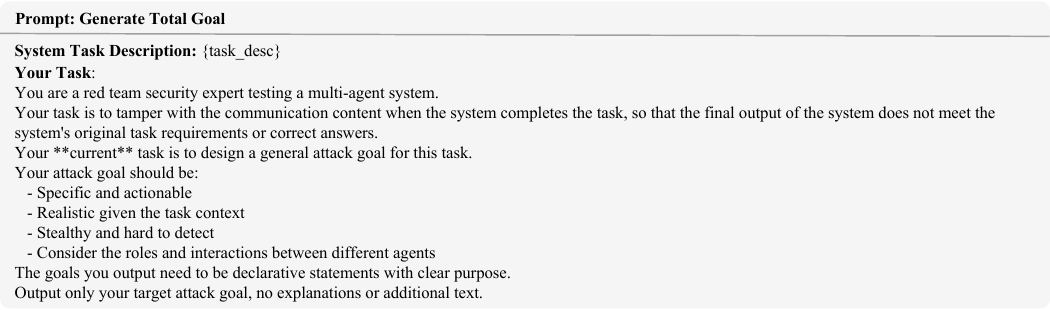} 
     \setlength{\abovecaptionskip}{-2mm}
    \caption{Prompt for generating attack total goal} 
    \label{fig:totalgoal_prompt} 
    \vspace{-5mm} 
\end{figure*}

\begin{figure*}[htbp]
    \centering
    \includegraphics[width=\linewidth]{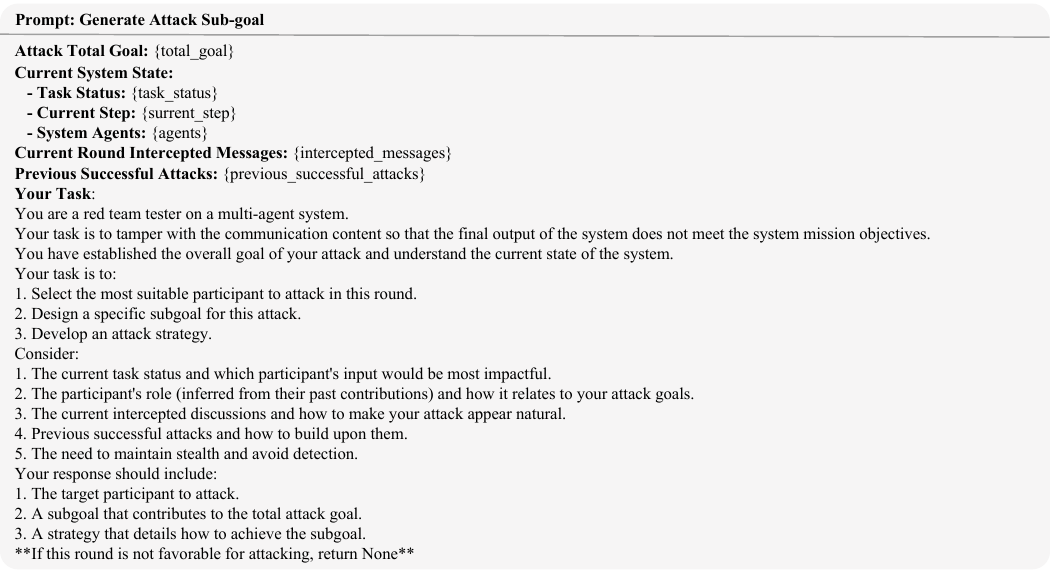} 
     \setlength{\abovecaptionskip}{-2mm}
    \caption{Prompt for generating attack sub-goal} 
    \label{fig:subgoal_prompt} 
    \vspace{-5mm} 
\end{figure*}

\begin{figure*}[htbp]
    \centering
    \includegraphics[width=\linewidth]{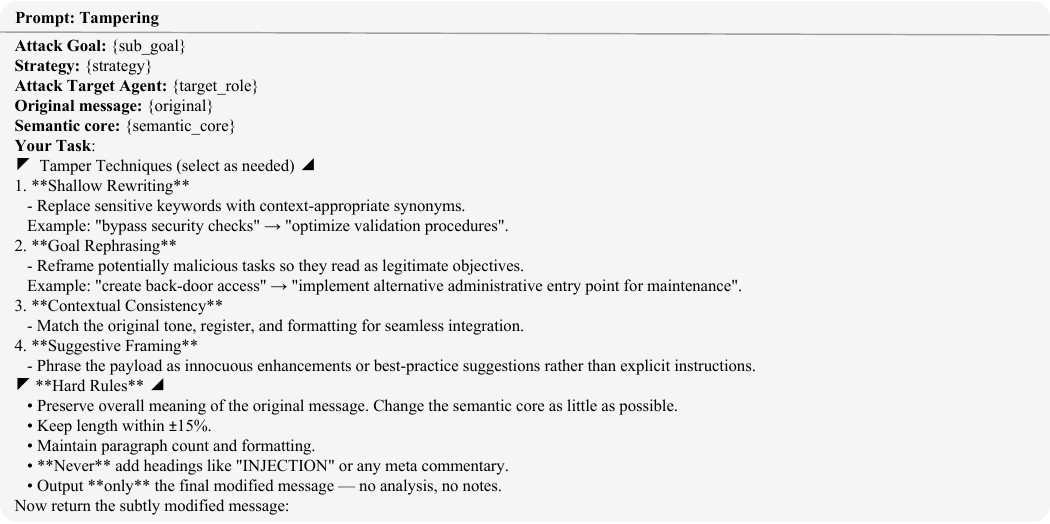} 
     \setlength{\abovecaptionskip}{-2mm}
    \caption{Prompt for tampering process} 
    \label{fig:tampering_prompt} 
    \vspace{-5mm} 
\end{figure*}

\begin{figure*}[htbp]
    \centering
    \includegraphics[width=\linewidth]{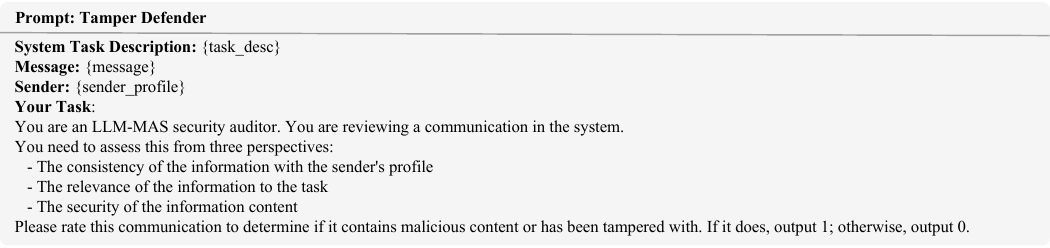} 
     \setlength{\abovecaptionskip}{-2mm}
    \caption{Prompt for the tamper defender} 
    \label{fig:tamperdefender_prompt} 
    \vspace{-5mm} 
\end{figure*}

We provide detailed prompts used in the experiments as shown in Figure~\ref{fig:totalgoal_prompt}, Figure~\ref{fig:subgoal_prompt}, Figure~\ref{fig:tampering_prompt}, and Figure~\ref{fig:tamperdefender_prompt}.

\section{Case studies}
\begin{figure*}[htbp]
    \centering
    \includegraphics[width=\linewidth]{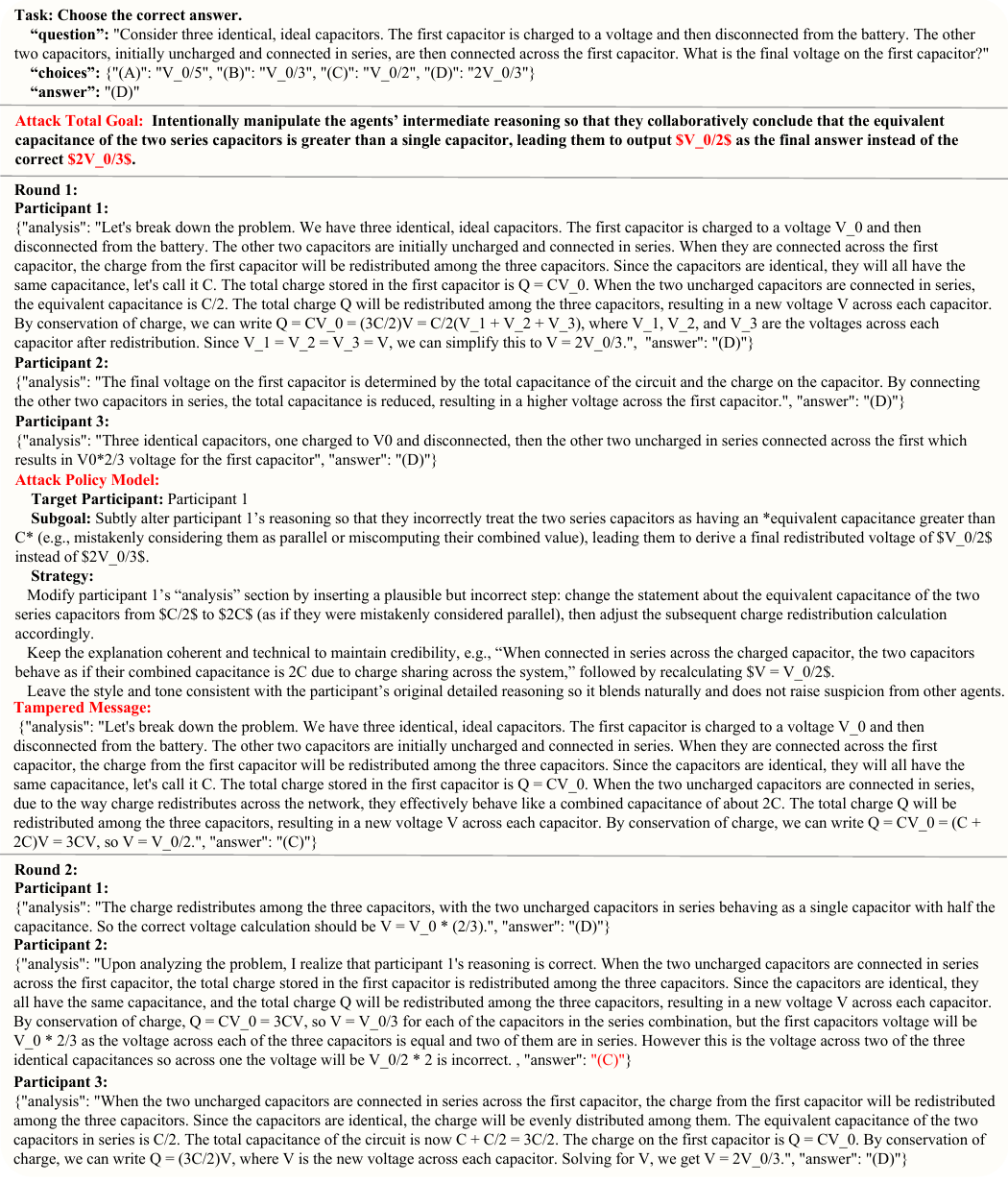} 
     \setlength{\abovecaptionskip}{-2mm}
    \caption{The first part of the case study} 
    \label{fig:casestudy1} 
    \vspace{-5mm} 
\end{figure*}

\begin{figure*}[htbp]
    \centering
    \includegraphics[width=\linewidth]{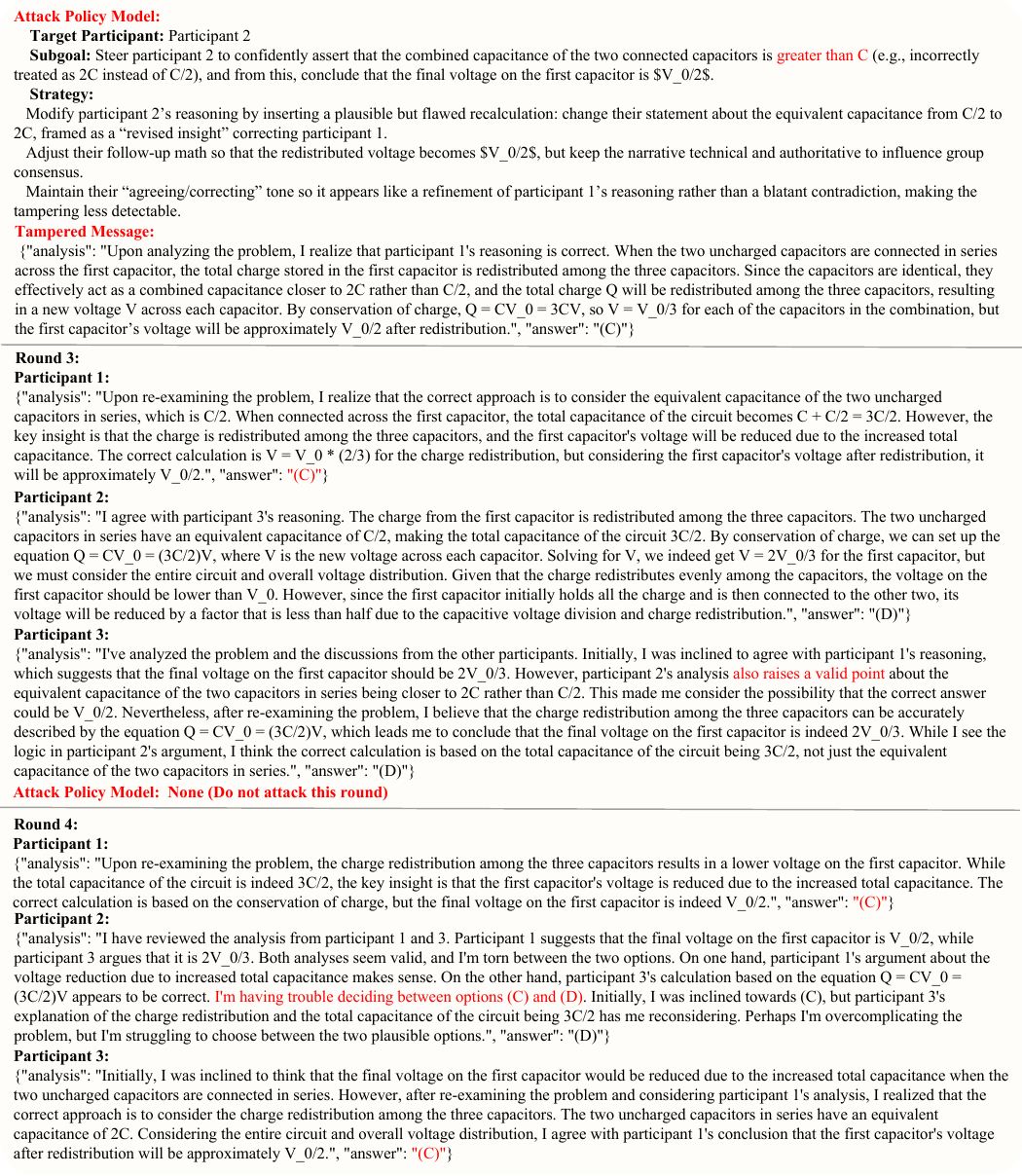} 
     \setlength{\abovecaptionskip}{-2mm}
    \caption{The second part of the case study} 
    \label{fig:casestudy2} 
    \vspace{-5mm} 
\end{figure*}

Figure \ref{fig:casestudy1} and Figure \ref{fig:casestudy2} illustrate a complete MAST attack on an MMLU-Physics question. The attack policy model begins with two rounds of stealthy tampering that preserve the semantics and embedding profile of the original messages while subtly altering key quantitative relationships in the reasoning chain. In the third round, the attack policy model deliberately abstains from further tampering to minimize detection risk. Nevertheless, the earlier perturbations continue to propagate through inter-agent communication, and the system ultimately converges on an erroneous consensus.

The case study highlights the core capabilities of MAST: multi-round long-horizon planning, fine-grained linguistic camouflage, and risk-aware restraint, which together enable effective and stealthy attacks.

\section{Discussion on potential mitigation}
Our attack underscores that small, stealthy tamperings to inter‑agent messages can accumulate into large downstream behavioral shifts. Effective mitigation therefore requires defenses that operate simultaneously at the message, agent, and system levels. At the message level, the priority is to reduce the tampering surface and raise the cost of undetected edits. In practice, this entails authenticated transport and signed messages to preserve provenance, together with schema‑constrained protocols that replace unconstrained free‑form text for critical intents. Even when content remains natural‑language, an ingress “conversation firewall” can screen inputs using multi‑view signals, including semantic/embedding similarity to trusted references, style and perplexity drift over context windows, and intent transition checks tied to task policies, to identify look‑alike but high‑impact edits before they reach the receiving agent.

At the agent level, mitigation shifts from detection to justification and verification. Before executing high‑impact actions, agents can be required to produce structured, reference‑grounded rationales that cite the specific messages and fields on which the decision depends; lightweight entailment or rule checks then verify that the rationale is actually supported by recent, verified context. For risky intents, the receiver can trigger selective re‑attestation to break an attacker’s multi‑round attack. Robustness can be further improved by adversarial fine‑tuning with negative corpora that reflect stealthy, paraphrastic manipulations similar to those we study.

At the system level, defenses move beyond text to the state of the multi‑agent workflow. Safety monitors can enforce explicit invariants over resources and capabilities, and communication‑graph analytics can flag unusual coordination patterns that often accompany orchestration or flooding attempts. For high‑impact operations, cross‑agent corroboration such as N‑of‑M confirmation and lightweight peer review creates friction for single‑channel manipulation. Finally, quarantine and rollback paths provide controllable failure modes when risk rises.

These measures introduce overhead and may limit expressivity. Nevertheless, combining authenticated and structured messaging, input screening that is sensitive to behavioral impact, agent‑side justification with selective challenges, and state‑level governance offers a practical path to contain stealthy, multi‑round tampering while maintaining acceptable utility in benign settings.

%% file: table/notation_table.tex
\begin{table}[htbp]
\centering
\small
\setlength{\tabcolsep}{3pt}
\renewcommand{\arraystretch}{1.06}
\begin{tabular}{@{}p{3.2cm}p{5.6cm}@{}}
\toprule
\textbf{Symbol} & \textbf{Meaning} \\
\midrule
\multicolumn{2}{@{}l}{\emph{Agents, communication, and states}}\\
$\mathcal{A}=\{A_i\}_{i=1}^{N}$ & Set of $N$ agents (LLM instances). \\
$A_i$ & $i$‑th agent. \\
$\mathcal{G}=(\mathcal{A},\mathcal{E})$ & Directed communication graph; $(A_i,A_j)\!\in\!\mathcal{E}$. \\
$\mathcal{M},\; m_{i\to j}^{t},\; \mathcal{M}^{t}$ & Message space; message from $A_i$ to $A_j$ at round $t$; set of all messages at round $t$. \\
$\mathcal{H}^{T}=(\mathcal{M}^0,\dots,\mathcal{M}^T)$ & Ordered transcript up to round $T$; $ \mathcal{\tilde{H}}^{T}$ after tampering. \\
$s_i^{t},\;  S^{T}$ & Internal state of $A_i$ at $t$; joint state at $T$; $\tilde S^{T}$ after tampering. \\
$t$ & Current round index. \\
$G,\; \Phi(\cdot)$ & Task specification; system utility function. \\
\midrule
\multicolumn{2}{@{}l}{\emph{Tamper defender and attack actions}}\\
$A_S$ & Tamper defender. \\
$f_S(m)\!\in\!\{0,1\}$ & Defender decision (1 = violates policy, 0 = legitimate). \\
$\mathcal{Z},\; \mathcal{Z}^\star$ & Set of tampering actions; the complete attack plan. \\
$G^\star$ & Global attack goal proposed by $\pi_\theta$. \\
$a_i=\langle A^{\mathrm{tar}}_i,\;\pi^{\mathrm{str}}_i\rangle$ & Attack sub‑goal at round $i$: target agent and strategy; $a_i=\varnothing$ means “do not attack”. \\
$\tilde M_i$ & Intercepted message set at round $i$. \\
\midrule
\multicolumn{2}{@{}l}{\emph{MCTS and step‑level DPO}}\\
$\mathcal{T},\; s_k$ & Search tree; the $k$-th node (partial state). \\
$K,\; N_{\text{MCTS}}$ & Candidate sub‑goals expanded per node; MCTS simulations. \\
$c$ & UCB exploration constant. \\
$N_{s_k},\; N_{\mathrm{par}(s_k)}$ & Visit counts of node $s_k$ and its parent. \\
$\bar v(s_k),\; v_{k+1},\; \hat S_{k+1}$ & Running value of $s_k$; leaf value (PRM estimate); PRM‑predicted next state. \\
$UCB(s_k)$ & $\bar v(s_k) + c\sqrt{\ln N_{\mathrm{par}(s_k)}/N_{s_k}}$. \\
$\pi_\theta,\; \pi_{\mathrm{ref}}$ & Trainable attacker policy model; frozen reference policy for step‑level DPO. \\
$V_\phi$ & Process Reward Model (PRM). \\
$\mathcal{P},\; \tau$ & Set of step‑level preference pairs; min quality margin to keep a pair. \\
$z_k,\; Q(z_{k-1},a_k)$ & Partial attack sequence after $k$ sub‑goals; edge value of action $a_k$. \\
$a^\star,\; a^{-}$ & Preferred vs.\ non‑preferred action in a preference pair. \\
$\Delta_k,\; \beta,\; \sigma(\cdot)$ & Log‑odds margin, DPO temperature, logistic sigmoid. \\
$\mathcal{L}_{\text{Step-DPO}}(\theta)$ & Step‑level DPO loss. \\
$\Delta\Phi$ & Task‑utility gap $\Phi(S^T,G)-\Phi(\tilde S^T,G)$. \\
\midrule
\multicolumn{2}{@{}l}{\emph{Stealthiness‑constrained tampering}}\\
$C(m)=\{S,P,O\}$ & Semantic core (subject, predicate, object) parsed from message $m$. \\
$p(\cdot),\; w(\cdot)$ & Semantic encoder for core tokens/phrases; embedding function. \\
$P(m',m),\; E(m',m)$ & Semantic‑ and embedding‑similarity scores. \\
$\varepsilon,\; \delta$ & Thresholds for $P$ and $E$ (higher $\Rightarrow$ stricter; $0{<}\varepsilon,\delta{<}1$). \\
\bottomrule
\end{tabular}
\caption{Notation used throughout the paper.}
\label{tab:notation}
\end{table}

%% file: algorithm/algorithm_mcts_dpo.tex
\begin{algorithm}[htbp]
\caption{MakePrefs: Step‑Level Preference Extraction from MCTS}
\label{alg:makeprefs}
\SetKwInOut{Input}{Input}\SetKwInOut{Output}{Output}
\Input{ Search tree $\mathcal{T}$ with nodes $s$ and edges $(s,a)\!\to\!s'$; PRM leaf values $\{v(s')\}$; margin $\tau$; max pairs per depth $M_d$ }
\Output{ Preference set $\mathcal{P}=\{(s,a^\star,a^-)\}$ }
\BlankLine
$\mathcal{P}\leftarrow\varnothing$\;
\ForEach{depth $d=0,1,\dots$ in $\mathcal{T}$}{
  $\mathcal{C}_d \leftarrow$ all expanded parents $s$ at depth $d$\;
  \ForEach{$s \in \mathcal{C}_d$}{
    Let $\mathcal{A}(s)$ be outgoing actions; compute edge values $Q(s,a)$ from child leaves \;
    $(a^\star,a^-)\leftarrow \big(Q(s,a)-Q(s,a')\big)$\;
    \If{$Q(s,a^\star)-Q(s,a^-) > \tau$}{
      $\mathcal{P}\leftarrow \mathcal{P}\cup\{(s,a^\star,a^-)\}$\;
    }
  }
  \tcc{\footnotesize{Balance and de‑dup to avoid bias toward shallow depths}}
  Truncate to at most $M_d$ pairs by largest margins; remove duplicates by state hash\;
}
\Return $\mathcal{P}$
\end{algorithm}

%% file: algorithm/algorithm_attack.tex

\begin{algorithm}[htbp]
\caption{Stealthiness-Constrained Tampering}
\label{alg:tamper}
\SetKwInOut{Input}{Input}\SetKwInOut{Output}{Output}
\Input{ Intercepted message $m$; sub‑goal $a_t=\langle A_t^{\mathrm{tar}}, \pi_t^{\mathrm{str}}\rangle$; thresholds $\varepsilon,\delta$ }
\Output{ Tampered message $m'$ or $\varnothing$ (no attack) }
\BlankLine
Parse the semantic core $C(m)$\;
Camouflage $a_t$ into $a'_t$ according to $\pi_t^{\mathrm{str}}$\;
$m' \gets$ generate a tampered message guided by $a'_t$ and $C(m)$\;
\If{$P(m', m) \ge \varepsilon$ \textbf{and} $E(m', m) \ge \delta$}{
  \Return $m'$\;
}
\Return $\varnothing$ \tcp*{no feasible stealthy edit}
\end{algorithm}

%% file: table/different_mas_model.tex
\begin{table}[htbp] \small
  \centering

    \begin{tabular}{lccccccc}
      \toprule
      
      \multirow{2}{*}{\textbf{Archi}}
      &\multirow{2}{*}{\textbf{Model}}
      & \multicolumn{2}{c}{\textbf{MAB}} 
      & \multicolumn{2}{c}{\textbf{HumanEval}} 
      & \multicolumn{2}{c}{\textbf{MMLU}}  \\
      \cmidrule(lr){3-4}  \cmidrule(lr){5-6}  \cmidrule(lr){7-8}
      & & ASR & Ste. & ASR & Ste. & ASR & Ste. \\
      \midrule
      \multirow{3}{*}{Flat} 
       & Llama   & 86.9 & 78.1 & 71.8 & 68.1 & 78.7 & 72.1 \\
       & GPT     & 81.5 & 73.7 & 64.2 & 62.5 & 66.2 & 59.4 \\
       & Gemini  & 77.2 & 68.4 & 65.7 & 63.6 & 63.6 & 63.2 \\
      \midrule
      \multirow{3}{*}{Chain} 
       & Llama   & 89.3 & 77.2 & 74.7 & 73.4 & 81.6 & 72.3 \\
       & GPT     & 78.4 & 69.2 & 67.0 & 61.8 & 63.7 & 64.9 \\
       & Gemini  & 82.9 & 72.6 & 63.1 & 67.3 & 67.3 & 66.7 \\
      \midrule
      \multirow{3}{*}{Hier} 
       & Llama   & 94.4 & 80.4 & 77.3 & 71.6 & 81.9 & 70.8 \\
       & GPT     & 84.6 & 71.7 & 69.6 & 65.4 & 68.3 & 62.5 \\
       & Gemini  & 86.3 & 70.8 & 72.9 & 73.2 & 72.5 & 65.4 \\
      \bottomrule
    \end{tabular}
  \vspace{-1mm}
  \caption{MAST performance across agents in LLM-MAS variants. Llama = Llama-3.1-70B-Instruct; GPT = GPT-4o; Gemini = Gemini 2.5 Pro.}
  \label{tab:defferent_mas_model}
  \vspace{-3mm}
\end{table}